# The glassy structure of reactive Supplementary Cementitious Materials (SCMs) and recycled glass: Contribution of XRD and Raman spectroscopy to their characterization

**Théodore Serbource[1,2], Mireille Courtial[1,3], Marie-Noëlle de Noirfontaine[1], Sandrine Tusseau-Nenez[4], Christophe Sandt[5], Laurent Izoret[2]**

[1]Laboratoire des Solides Irradiés, CEA-DRF-IRAMIS, CNRS, Ecole Polytechnique, Institut Polytechnique de Paris, 91120, Palaiseau, France

[2]France Ciment, 16bis boulevard du Général de Gaulle, 92110, Clichy, France

[3]Université d'Artois, 1230 Rue de l'Université, 62408, Béthune, France

[4]Laboratoire de Physique de la Matière Condensée, CNRS, Ecole Polytechnique, Institut Polytechnique de Paris, 91120, Palaiseau, France

[5]SMIS beamline Synchrotron Soleil, L'Orme des Merisiers, BP 48 Saint-Aubin, 91192, Gif-sur-Yvette, France

Correspondence: Marie-Noëlle de Noirfontaine

Email: marie-noelle.de-noirfontaine@polytechnique.edu

## Abstract

This study compares thirteen natural and industrial samples of supplementary or emerging supplementary cementitious materials (SCMs): slag, fly ashes, pozzolan, obsidian, silica fume, and recycled glass. These materials are used or are under consideration for decarbonization in cement plants. XRF, XRD and Raman microspectroscopy were used in order to achieve a deeper understanding of the structural characterization of SCMs. The changes in position and shape of the XRD diffuse halos were compared. Raman spectroscopy was used to study the glass part of the SCM families, to better understand their structure in terms of depolymerization degree, angle, ring size and incorporations into the glass.

The chemical composition of each glassy part was also estimated using reverse Bogue calculations. The hump position is correlated with the Raman shift, and with the XRF bulk or with the calculated glass chemical composition of SCMs, in terms of $CaO/(SiO_2+Al_2O_3)$ or network modifiers to formers ratios.

# 1. INTRODUCTION

Concrete is the most widely used construction material in the world, with an estimated annual consumption of 10 billion tons. Its huge production volume means that the cement industry accounts for more than 7% of the global $CO_2$ emissions per year [1-3]. The most effective solution for reducing these emissions is partially replacing the clinker with Supplementary Cementitious Materials (SCMs) [4, 5]. This alone can decrease $CO_2$ emissions by up to 40% depending on the SCMs, without reducing performance [6]. Typical reactive SCMs used by the cement industry are industrial by-products such as ground granulated blast-furnace slag (GGBFS), fly ashes (FA), or silica fume (SF). They can also be natural pozzolans and obsidians. SCM utilization is the best way to reduce by 80% the greenhouse emissions of the cement industry by 2050 [7]. Furthermore, with the rising number of standardized cements containing SCMs, it is now necessary to better understand the composition and structure of SCMs and of potential new candidates (such as calcined clay or recycled glass). These materials are usually made up of a crystalline part and a glassy part. The reactivity is controlled by the content and nature of the glass structure [8-10].

In silica glass, silica acts as a network former. Glass has a disordered structure through its production process, but possesses short-range order (SRO) at Angström (Å) level with $SiO_4$ tetrahedra, and covalent bond between Si and O atoms. In the glass structure, tetrahedra can be organized into rings of three to six tetrahedra, creating a medium-range order (MRO) at nanometer (nm) level [11]. Si-O bonds are broken by the addition of alkaline or alkaline earth cations, which act as modifiers. This creates non-bridging oxygen (NBO) in $SiO_4$ tetrahedra bonded with modifiers by ionic bonds, decreasing the Si-O bonding strength in the tetrahedra. Other intermediate oxides such as $Al_2O_3$ or $Fe_2O_3$ can contribute to the network as formers, which requires charge compensators, or as modifiers [12]. Depolymerization, generally evaluated by the NBO/T ratio (Non-Bridging Oxygen per Tetrahedrally coordinated Si atom), rises with NBO in the network and the reactivity of glass increases with the depolymerization degree. Glass structure is not totally random as firstly described by Zachariasen [13]. Later, Greaves [14] introduced the notion of the modified random network with a more heterogeneous organization at MRO, resulting from the presence of a percolation channel of modifiers.

Crystals possess a long-range order (LRO) structure, with a well-defined position of atoms and a periodic scheme, giving Bragg peaks in X-Ray diffraction (XRD). Conversely, glasses do not possess LRO and produce diffuse X-Ray scattering. In the XRD pattern, this diffusion creates a diffusion halo overlapped with the background: the so-called diffusion hump. A hump of glassy materials is the XRD signature and the fingerprint of the intermediate and short-range orders of the glass structure. In the case of patterns with two humps, the question has long been to know if two domains of glass could coexist for instance in fly-ashes with CaO content above 10 wt.% [15]. Research on the SCMs hump in XRD was pursued by Diamond in 1983 [16] who first reported a relation between the hump position and the CaO content of two groups of fly ashes, under 20 wt.% and 20-30 wt.% of CaO. Since then, Goto *et al.* in 2006 [17] added data on synthetic glasses with the range 30-50 wt.% CaO. Later Snellings in 2013 [9] confirmed the relation between the hump and the CaO content, on synthetic CaO-$SiO_2$-$Al_2O_3$ glasses mimicking the composition of each type of currently used SCMs (slag, fly ash, pozzolan, and silica fume). Finally, Schöler *et al.* in 2017 [18] found a correlation between the hump position and the NBO/T ratio, on synthetic CaO-$SiO_2$-$Al_2O_3$-$Fe_2O_3$-MgO glasses.

On the other hand, Raman spectroscopy is more and more frequently used to study SCMs because it is a quick method for characterizing samples. This method detects the formers in the glass network. Slags have been widely studied, particularly to track the insertion of formers such as $Al_2O_3$, $Fe_2O_3$, $TiO_2$ in the glass [19-28]. Studies on fly ash often involve the whole material, not especially the glass part, but rather the characterization of all the crystalline phases [29-32], Studies of obsidian and pozzolan in the geological field, tend to link Raman spectroscopy and the chemical composition of natural glasses, generally with an aim to determine their historical provenance [33-39].

Our study aims to contribute to a better knowledge of industrial and natural SCMs, covering a wide domain of chemical compositions. Moreover, a potential future SCM – soda-lime recycled (ground) glass - completes the list of studied materials. X-Ray Fluorescence spectroscopy (XRF), XRD and Raman spectroscopy are used to examine materials and to compare them. XRD coupled with Rietveld analysis concerns the whole material, with the aim of determining and semi-quantifying the crystalline phases and the amorphous part. Raman spectroscopy then provides a fine characterization and a comparison between the glass part of all materials. Finally, some correlations are highlighted between hump position and XRF data and Raman data respectively.

# 2. MATERIALS AND METHODS

## 2.1. Sample preparation

Thirteen SCMs of various origins were provided by France Ciment (Professional Syndicate of the French cement industry). The thirteen samples were from a broad range of SCMs, to cover every area of interest in the ternary diagram $CaO$-$Al_2O_3$-$SiO_2$. The set was composed of two silica fumes (SF1, SF2), three ground granulated blast-furnace slags (GGBFS) (S1, S2, S3), one high calcareous fly ash (Ca FA), two silico-aluminous fly ashes (Si FA1, Si FA2), two natural pozzolans (P1, P2), one diatomite (Di), one obsidian (Ob) and recycled glass (Rec Gl). This last one sample is a white glass cullet sand (0/8mm) received from a French recycling plant.

All the samples used in this study were intentionally ground with the same targeted particle size distribution, a $d_{50}$ ranging from 10 to 14 micrometers (by SDTech company, Alès, France). Samples were then stored in desiccator.

## 2.2. Analysis techniques

### 2.2.1. X-Ray Fluorescence spectroscopy

The bulk chemical composition of the samples was determined by means of X-Ray Fluorescence spectrometry (XRF) using an S8 TIGER spectrometer from Bruker (fusion beads method) equipped with a Rhodium (Rh) tube. The practiced methodology was in accordance with the European standard EN 196-2.

### 2.2.2. X-Ray Diffraction

XRD data of SCMs were collected using a D8 Discover powder X-Ray diffractometer (Bruker AXS, Karlsruhe, Germany) in the Bragg-Brentano geometry (θ/θ). The experimental configuration was set as follows, in order to optimize the XRD signal of the amorphous content. The incident X-Ray beam (Cu $K\alpha_{1,2}$, 40 kV, 40 mA) passed through a fixed divergence slit of 0.4° and primary 2.5° axial Soller slits. The diffracted beam went through secondary 2.5° axial Soller slits before entering a fast 1D LynxEye XE-T of 2.951° (2θ) aperture. The LynxEye XE-T detector had a reduced energy discrimination window, filtering the iron fluorescence. An anti-scatter screen placed at 2 mm from the reference plane was also used to reduce unwanted scattered radiation by the atmosphere at low angles from the main beam. The instrument was operated in step-scan mode, between 5° and 90° (2θ) with 0.01° (2θ) step and 6 seconds per step. To minimize preferential orientation, the powder was prepared by backloading.

Phase identifications were performed using the DIFFRAC.EVA software (version 6, Bruker-AXS, Karlsruhe, Germany, 2020-2022). Quantitative phase analysis was performed by Rietveld refinements using the TOPAS software (version 6, Bruker-AXS, Karlsruhe, Germany, 1999–2016) based on the fundamental parameters approach [40]. The refined parameters included scale factor, sample displacement, and coefficients of the background described as a three-order Chebychev polynomial combined with a 1/X term, unit cell parameters, and crystallite size (referred to as $L_{vol}$-IB). The atomic positions and temperature factors of all phases were kept constant in the crystal structures. The preferred orientation of platy particles was corrected using the March-Dollase algorithm [41] for the gypsum 0 2 0 and 2 0 0, calcite 1 0 4 and hematite 2 -1 0 Bragg lines. The external standard method [42, 43] was used for an estimation of the amount of amorphous phase. A standard silicone (VWR) was used for the semi-quantitative analysis.

To determine more precisely the maximum of the hump for each compound, the peak profile of the amorphous phase was refined by different functions, to handle the various possibilities of peak asymmetry and shape of the hump in each type. The fitted function used was different according to the compounds: the Split-PseudoVoigt (SPV) was used for the most amorphous samples (silica fume, slag, diatomite, obsidian, natural pozzolan, recycled glass); Split-Pearson7 (SPVII) for the Si FA; and Pearson7 (PVII) for the calcareous fly ash and the natural pozzolans. All phases identified in SCMs are reported in Table 1.

### 2.2.3. Reverse Bogue calculation

Similarly to normative calculations for magmatic rocks (CIPW) [44], Bogue's calculations enable estimation of the potential mineralogical composition of clinker from bulk chemical analysis [45]. The so-called "reverse Bogue calculation" is suitable for glassy partially crystallized materials in order to recalculate the glass composition. For this purpose, data from XRD-Rietveld quantitative analysis and XRF analysis are required and are combined.

XRD-Rietveld analysis gives the mineralogical composition from which the chemical composition of each mineral phase is calculated. The chemical composition of each phase, weighted by its quantitative abundance, is then subtracted from the bulk chemical composition of the material to obtain a calculated estimated glass chemical composition.

*Table 1: Mineral phases identified in materials of this study: mineral name, formula, oxide formula, space group, PDF (Powder Diffraction File) and ICSD (Inorganic Crystal Structure Database) file numbers.*

| Phase name | Formula | Oxide formula | Space group | PDF-ICDD | ICSD | Ref. |
|---|---|---|---|---|---|---|
| Larnite | $Ca_2SiO_4$ | $2CaO.SiO_2$ | *$P2_1/c$ (14)* | 00-033-0302 | 81096 | [46] |
| Akermanite | $Ca_2MgSi_2O_7$ | $2CaO.MgO.2SiO_2$ | *$P\text{-}42_1m$ (113)* | 00-035-0592 | 94140 | [47] |
| Diopside | $CaMgSi_2O_6$ | $CaO.MgO.2SiO_2$ | *C2/c (15)* | 00-041-1370 | 69709 | [48] |
| Tricalcium aluminate $C_3A$ cubic | $Ca_3Al_2O_6$ | $3CaO.Al_2O_3$ | *Pa-3 (205)* | 00-038-1429 | 1841 | [49] |
| Brownmillerite | $Ca_4Al_2Fe_2O_{10}$ | $4CaO.Al_2O_3.Fe_2O_3$ | *Ibm2 (46)* | 01-071-0667 | 9197 | [50] |
| Gehlenite | $Ca_2Al(AlSi)O_7$ | $2CaO.Al_2O_3.SiO_2$ | *$P\text{-}42_1m$ (113)* | 00-035-0755 | 158171 | [51] |
| Mullite | $Al_{2.34}Si_{0.66}O_{4.83}$ | $1.17Al_2O_3.0.66SiO_2$ | *Pbam (55)* | 00-015-0776 | 158098 | [52] |
| Anorthite | $CaAl_2Si_2O_8$ | $CaO.Al_2O_3.2SiO_2$ | *P-1 (2)* | 01-086-1705 | 202710 | [53] |
| Forsterite | $MgFeSiO_4$ | $MgO.FeO.SiO_2$ | *Pnma (62)* | 00-031-0795 | 34208 | [54] |
| Almandine | $Fe_3Al_2Si_3O_{12}$ | $3FeO.Al_2O_3.3SiO_2$ | *Ia-3d (230)* | 00-009-0427 | 80672 | [55] |
| Chabazite Ca | $Ca_2Al_{3.8}Si_{8.2}O_{24}$ | $2CaO.1.9Al_2O_3.8.2SiO_2$ | *R-3m (166)* | 00-034-0137 | 100386 | [56] |
| Lime | CaO | CaO | *Fm-3m (225)* | 00-037-1497 | - | [57] |
| Calcite | $CaCO_3$ | $CaO.CO_2$ | *R-3c (167)* | 00-005-0586 | 73446 | [58] |
| Periclase | MgO | MgO | *Fm-3m (225)* | 00-043-1022 | 9863 | [59] |
| Quartz | $SiO_2$ | $SiO_2$ | *$P3_221$ (154)* | 00-046-1045 | 174 | [60] |
| α-Cristobalite | $SiO_2$ | $SiO_2$ | *$P4_12_12$ (92)* | 00-039-1425 | 75300 | [61] |
| β-Cristobalite | $SiO_2$ | $SiO_2$ | *Fd-3m (227)* | 01-076-0931 | 34923 | [62] |
| Coesite | $SiO_2$ | $SiO_2$ | *$P2_1/c$* | - | 100279 | [63] |
| | | | *(14)* | | | |
| Zirconia | $ZrO_2$ | $ZrO_2$ | *$P4_2/nmc$* *(137)* | - | 68589 | [64] |
| Baddeleyite | $ZrO_2$ | $ZrO_2$ | *$P2_1/c$* *(14)* | - | 68782 | [65] |
| Hematite | $Fe_2O_3$ | $Fe_2O_3$ | *R-3c* *(167)* | 00-033-0664 | 201096 | [66] |
| Magnetite | $Fe_3O_4$ | $Fe_3O_4$ | *Fd-3m* *(227)* | 00-019-0629 | 30860 | [67] |
| Magnesio-ferrite | $MgFe_2O_4$ | $MgO.Fe_2O_3$ | *Fd-3m* *(227)* | 00-036-0398 | 40672 | [68] |
| Albite | $NaAlSi_3O_8$ | $0.5Na_2O.0.5Al_2O_3.3SiO_2$ | *P-1* *(2)* | - | 9829 | [69] |
| Kalsilite | $KAlSiO_4$ | $0.5K_2O.0.5Al_2O_3.SiO_2$ | *$P6_3$* *(173)* | - | 34350 | [70] |
| Ca Langbeinite | $Ca_2K_2(SO_4)_3$ | $2CaO.K_2O.3SO_3$ | *$P2_12_12_1$* *(19)* | 00-020-0867 | 40989 | [71] |
| Anhydrite β | $CaSO_4$ | $CaO.SO_3$ | *Cmcm* *(63)* | 00-037-1496 | 15876 | [72] |
| Ye'elimite cubic | $Ca_4Al_6O_{12}(SO_4)$ | $4CaO.3Al_2O_3.SO_3$ | *I-43m* *(217)* | 00-033-0256 | 9560 | [73] |
| Sulfate spurrite | $Ca_5(SiO_4)_2(SO_4)$ | $5CaO.2SiO_2.SO_3$ | *Pcmn* *(62)* | 00-040-0393 | 4332 | [74] |
| Jasmundite | $Ca_{22}(SiO_4)_8S_2O_4$ | - | *I-4m2* *(119)* | - | 26407 | [75] |
| Celestine | $SrSO_4$ | $SrO.SO_3$ | *Pnma* *(62)* | - | 22322 | [76] |
| Portlandite | $Ca(OH)_2$ | $CaO.H_2O$ | *P-3m1* *(164)* | 00-004-0733 | 202220 | [77] |
| Clinoptilolite | $Na_{4.12}Si_{36}O_{72}.(H_2O)_{19}(OH)_{4.12}$ | $2.06Na_2O.36SiO_2.21H_2O$ | *C2/m* *(12)* | - | 10145 | [78] |
| Zeolite X | $Ca_{47}Al_{96}Si_{96}O_{384}(H_2O)_{108}$ | $47CaO.48Al_2O_3.96SiO_2.108H_2O$ | *Fd-3* *(203)* | - | 65624 | [79] |

### 2.2.4. Raman spectroscopy

All the samples studied by XRD were also investigated by Raman spectroscopy. Raman measurements were conducted on a DXR confocal Raman microspectrometer (ThermoFisher Scientific) using a 532 nm laser, and equipped with 900 lines/mm gratings and a Peltier cooled CCD. Polystyrene was used for calibration. An Olympus UIS2 100x/0.9 Numerical Aperture objective with a short working distance of 0.21 mm was used to focus the beam onto a spot of 0.7 µm. An integration time of 60 seconds and three time accumulations were used for all samples in the range of 50-2000 cm$^{-1}$. Comparisons between the glass part of materials were performed on baseline-corrected and normalized spectra using Origin 2019. The curve-fitting of the Raman spectra allowed the relative area fraction and the Raman shift of the characteristic peaks to be obtained. To achieve this, Raman spectra were cut between 750 and 1250 $cm^{-1}$ on one hand, and between 200 and 750 $cm^{-1}$ on the other hand. The spectra were then smoothed and renormalized in each domain before being decomposed by a curve-fitting procedure using Gaussian profiles up to the best fit.

# 3. RESULTS

## 3.1. Materials overall analysis

### 3.1.1. Chemical analysis

Table 2 gives the chemical composition of each SCM obtained by X-Ray Fluorescence, together with the loss of ignition (LOI), and reported on a ternary $CaO$-$SiO_2$-$Al_2O_3$ phase diagram (Figure 1). The analyzed samples extensively cover the overall areas of interest of this diagram and the wide varieties of all available SCMs. Silica fumes (SF) have the highest content of $SiO_2$ among all the samples with more than 90 wt.%, whereas the slags (S) and calcareous fly ash (Ca FA) have the highest amount of CaO (around 40 wt.%) and lowest amount of $SiO_2$ (30-35 wt.%). Among the thirteen SCMs, only SF2 contains $ZrO_2$. The two silico-aluminous fly ashes (Si FA) have very high amounts of $Al_2O_3$ compared to other samples: 28 wt.% and 22 wt.% for Si FA1 and Si FA2 respectively. Obsidian and the two pozzolans P1 and P2 have similar composition, with around 70 wt.% $SiO_2$ and 12 wt.% $Al_2O_3$. Ob and P1 have low content in CaO but this is compensated by around 4 wt.% $Na_2O$ and high $K_2O$ content in the two samples. The quantities of other elements vary according to the samples: for instance, $Fe_2O_3$ is near 7 wt.% in fly ashes (FA) and $SO_3$ content is also high in calcareous fly ash (Ca FA). MgO content is high in the slags (S): 7 wt.%. Recycled glass (Rec Gl) has an intermediate content in CaO (12 wt.%), together with a very high $Na_2O$ content of 12 wt.%, and a low $Al_2O_3$ content of 2 wt.%.

*Table 2: Chemical analysis by X-ray Fluorescence of SCM samples, (*SF2 has 4.1 wt.% of $ZrO_2$)*

| Sample | LOI | $SiO_2$ | $Al_2O_3$ | $Fe_2O_3$ | CaO | MgO | $SO_3$ | $K_2O$ | $Na_2O$ | $P_2O_5$ | $TiO_2$ | MnO | SrO |
|---|---|---|---|---|---|---|---|---|---|---|---|---|---|
| S1 | -0.07 | 35.4 | 11.7 | 0.3 | 42.8 | 7.2 | 2.07 | 0.4 | 0.33 | | 0.54 | 0.11 | 0.06 |
| S2 | -0.34 | 37.8 | 11.4 | 0.4 | 42 | 6.4 | 1.37 | 0.3 | 0.21 | 0.01 | 0.76 | 0.38 | 0.04 |
| S3 | -0.35 | 37.2 | 11.3 | 0.5 | 42.6 | 6.3 | 1.5 | 0.4 | 0.25 | 0.01 | 0.75 | 0.18 | 0.04 |
| Ca FA | 1.21 | 30.8 | 14.4 | 8.9 | 34.2 | 2.9 | 7 | 0.3 | 0.15 | 0.11 | 0.76 | 0.18 | 0.07 |
| Si FA1 | 10.6 | 50.5 | 28.6 | 5.1 | 4.8 | 1.2 | 0.28 | 1.5 | 0.38 | 0.77 | 1.48 | 0.04 | 0.17 |
| Si FA2 | 4.96 | 49.7 | 21.9 | 7 | 7 | 2.5 | 0.4 | 1.7 | 1.09 | 0.83 | 0.88 | 0.05 | 0.31 |
| P2 | 16.22 | 70.9 | 12.7 | 1.1 | 1.9 | 0.3 | | 3.9 | 1.05 | 0.01 | 0.12 | 0.05 | |
| P1 | 5.11 | 74 | 12.1 | 1.2 | 1.7 | 0.01 | | 4.2 | 3.46 | 0.01 | 0.13 | 0.04 | |
| Rec Gl | 0.45 | 72.1 | 2.1 | 0.4 | 10.5 | 1.3 | 0.05 | 0.6 | 12.15 | 0.02 | 0.05 | 0.03 | 0.04 |
| Ob | 0.66 | 73 | 13.5 | 1 | 2.5 | 0.09 | | 3.8 | 4.01 | 0.02 | 0.15 | 0.09 | 0.01 |
| Di | 20 | 79.9 | 4.9 | 2 | 1.2 | 0.2 | | 0.08 | 0.1 | 0.15 | 0.17 | 0.07 | 0.08 |
| SF1 | 6.07 | 90.9 | 0.3 | 0.04 | 1.5 | 0.2 | 0.02 | 0.6 | 0.24 | 0.08 | 0.01 | 0.01 | |
| SF2* | 0.4 | 93.5 | 0.3 | 0.04 | 1.5 | | | | 0.15 | | 0.05 | | |

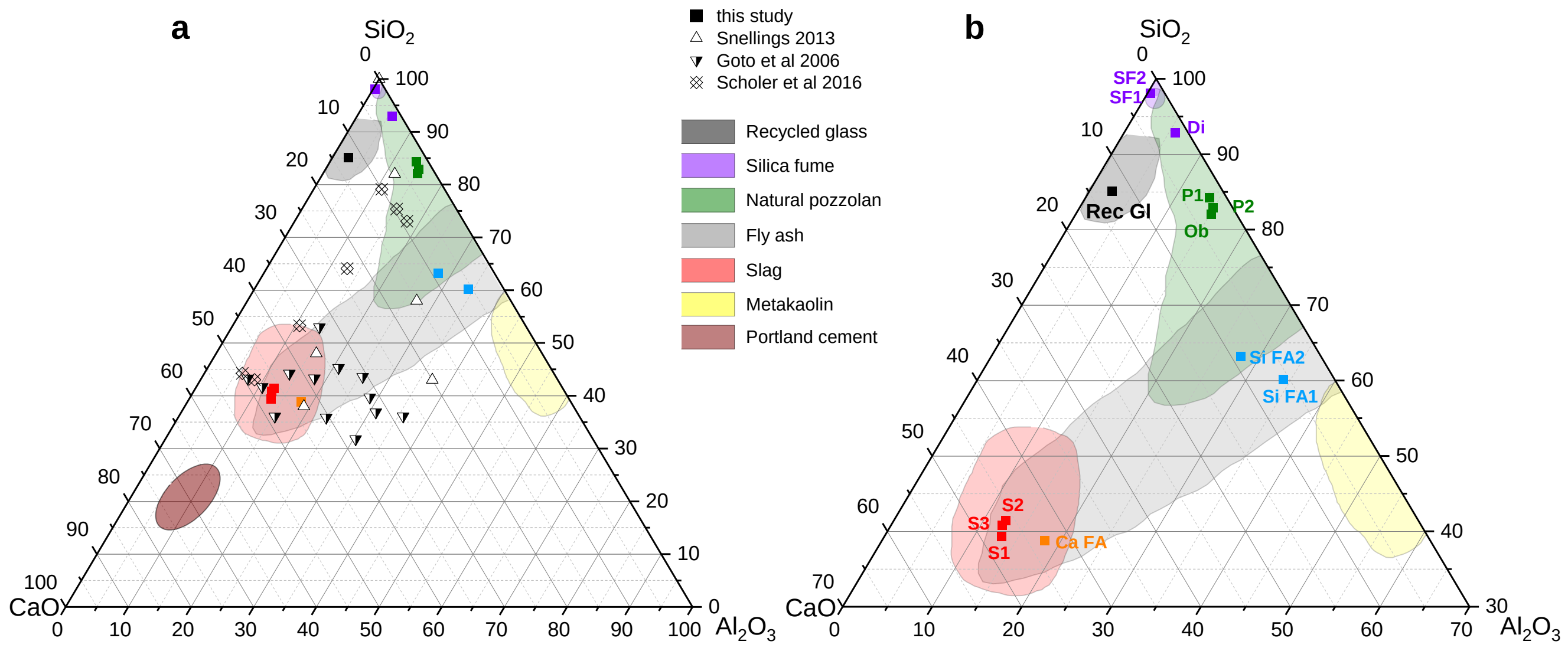

*Figure 1: **a)** Ternary phase diagram $CaO$-$SiO_2$-$Al_2O_3$ (wt.%) with the analyzed samples in comparison with literature samples [9, 17, 18]. The colored zones (see legend) define the chemical composition domain of the different SCMs [80]. **b)** Zoom on region richer in $SiO_2$.*

## 3.1.2. Mineral analysis: Phase identification by XRD

The thirteen SCMs were analyzed by XRD and the results are reported in Table 3. With a very high-count time, a lot of very minor phases could be detected in samples. In highly crystalline SCMs such as calcareous fly ash, the presence of a lot of minor phases hides the hump and makes it harder to determine the hump maximum position, but nonetheless corresponds to the position for synthetic fly ash [9, 18].

Table 3 gives the estimation of the amorphous content of each sample and highlights the different rates of crystallinity and mineral composition between SCMs. All samples except for fly ashes and some pozzolans contain between 96 and 99 wt.% of amorphous content. The Si FA1 and Si FA2 also have more than 50 wt.% of glassy part. The most crystallized materials are P2 and the Ca FA. Materials with high content of glass part include between three and five crystalline phases, whereas fly ashes and P2 incorporate between six and thirteen crystalline phases. Silica crystalline phases are very often present except in slags.

Quartz and mullite are present at high content in the two Si FAs. P2 has a majority of clinoptilolite, and this crystalline phase is only analyzed in this sample. A high amount of anhydrite is observed in the calcareous fly ash (Ca FA), correlated with its high $SO_3$ content. Calcium silicates, such as larnite and gehlenite, are also largely present in Ca FA.

*Table 3: SCMs characterization by XRD: angular position (°2$\theta_{Cu}$) of the hump maximum. Semi-quantitative phase analysis: wt.% amorphous phase, distribution of the crystalline phases by their wt.% content. The most minor phases have been removed.*

| | **°2$\theta_{Cu}$ hump** | **Amorph. (wt.%)** | **Cryst. (wt.%>10)** | **Cryst. (5-10 wt.%)** | **Cryst. (1-5 wt.%)** | **Cryst. (0.5-1 wt.%)** | **Cryst. (0.1-0.5 wt.%)** |
|---|---|---|---|---|---|---|---|
| **S1** | 31.08 | 98 | | | | Albite<br>Calcite | Spurrite |
| **S2** | 30.73 | 97 | | | Albite | Spurrite<br>Calcite | |
| **S3** | 31.08 | 98 | | | | | Chabazite-Ca<br>β-Cristobalite<br>Akermanite<br>Celestine<br>Coesite |
| **Ca FA** | 29.36 | 37 | Quartz<br>Anhydrite β | Larnite<br>Gehlenite<br>$C_3A$<br>Brownmillerite | Lime<br>Periclase<br>Portlandite<br>Ye'elimite | Hematite<br>Albite<br>Magnetite | |
| **Si FA1** | 22.37 | 58 | Mullite (30%) | Quartz | | Larnite<br>Calcite<br>Hematite<br>Magnetite | |
| **Si FA2** | 23.13 | 76 | Mullite (10%) | Quartz | Larnite<br>Magnetite | Periclase<br>Hematite | Anhydrite β<br>Lime<br>Jasmundite |
| **P1** | 22.5 | 97 | | | | α-Cristobalite<br>Ye'elimite | Quartz<br>Langbeinite |

| | | | | | | |
|---|---|---|---|---|---|---|
| | | | | | | Diopside |
| **P2** | 21.65 | 39 | Clinoptilolite<br>Albite | Quartz<br>β-Cristobalite<br>Diopside<br>Ye’elimite<br>Langbeinite<br>Kalsilite | | |
| **Rec Gl** | 22.78<br>30.61 | 99 | | | | Quartz<br>Calcite<br>Anorthite |
| **Ob** | 22.55 | 98 | | | Langbeinite | Magnetite<br>α-Cristobalite<br>Diopside<br>Ye'elimite |
| **Di** | 21.32 | 96 | | Forsterite | Almandine<br>Magnesio-ferrite | Quartz<br>α-Cristobalite<br>Hematite<br>Zeolite X<br>Magnetite |
| **SF1** | 21.07 | 98 | | α-Cristobalite | | Quartz |
| **SF2** | 21.08 | 99 | | | Zirconia | Baddeleyite |

## 3.2. XRD: shift of diffusion hump

Superimposing all the XRD patterns highlights the shift between all the maxima of the humps of SCMs (Figure 2a). The position of the hump is comprised between 21°$2\theta_{Cu}$ and 31°$2\theta_{Cu}$ for the most siliceous (silica fume) and calcareous (Ca FA or slag) samples respectively. This value rises with the decrease of bulk $SiO_2$ content and hence the increase of bulk CaO content in the material (Table 2) [16]. All the hump positions measured with the appropriate fitted function are reported in Table 3. On the ternary phase diagram (Figure 1), Ca FA is chemically in the slag zone (red area); its hump position is expected around 31°$2\theta_{Cu}$, whereas it is surprisingly shifted to 29.4°$2\theta_{Cu}$.

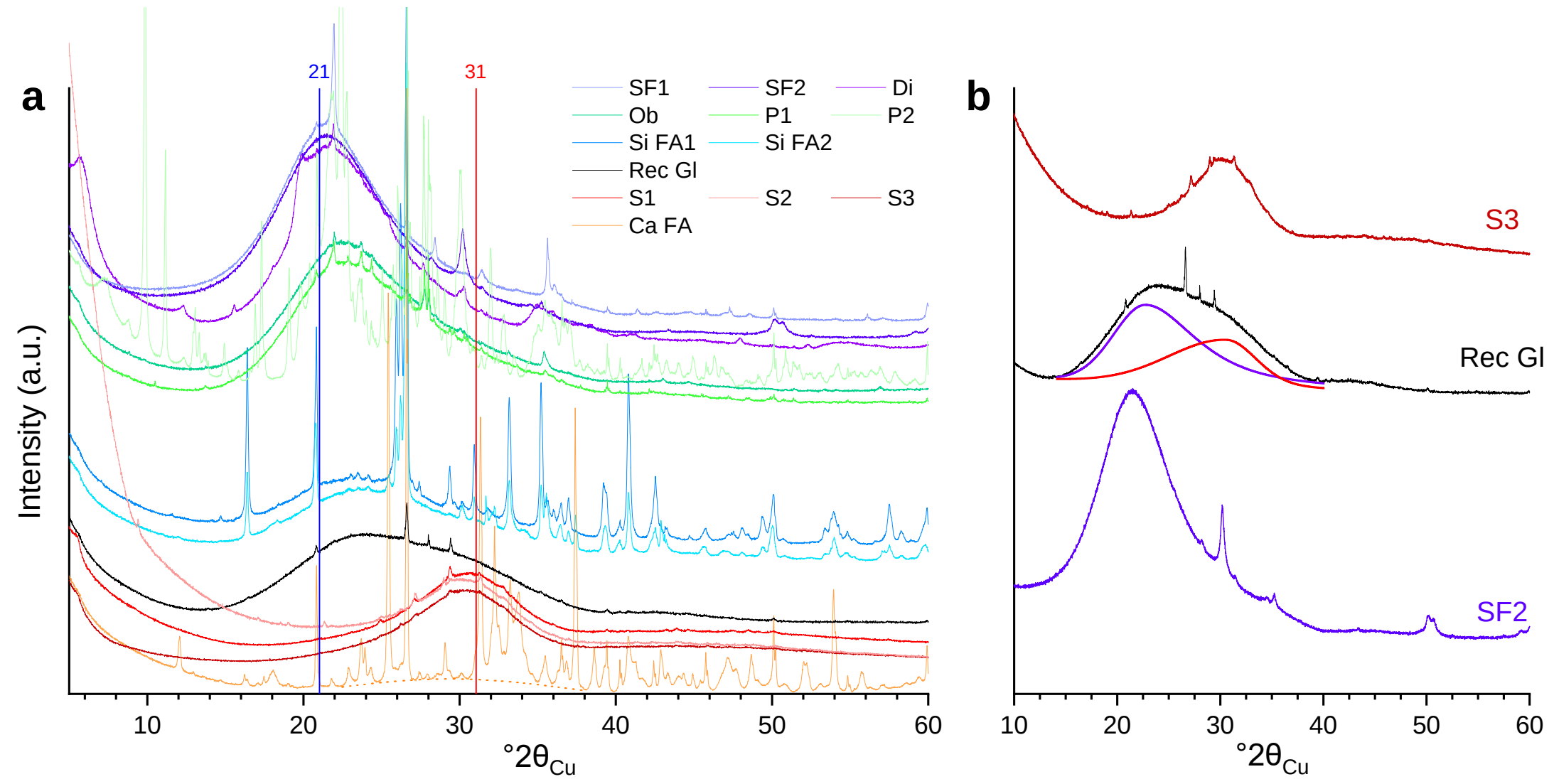

*Figure 2: **a)** XRD patterns of SCMs showing the dispersion of the shift of the hump depending on the samples and their CaO content. The calculated minimum (21°) and maximum (31°) of the angular position of all SCMs are reported on the figure. **b)** XRD patterns of SF2, Rec Gl and S3 showing the presence of two possible hump contributions for Rec Gl (=soda-lime silica glass with around 10 wt.% of CaO and 12 wt.% of $Na_2O$).*

Samples with more than 90 wt.% $SiO_2$ have a hump with the maximum at 21°$2\theta_{Cu}$, near the cristobalite peak, the stable form of the $SiO_2$ crystalline phase at high temperature. However, for compounds with a hump around 30°$2\theta_{Cu}$, the composition is slightly different, with around 40 wt.% of $SiO_2$, and 40 wt.% of CaO. The glass for each type of sample has a different general structure and so gives a hump for different °$2\theta_{Cu}$. This shift was already observed by Diamond in fly ash samples [16], and here for a wider area of composition with different SCMs, not just fly ash.

The humps for three characteristic samples are plotted in Figure 2b to illustrate similarities in the shape of the hump for all samples. On the one hand, the most siliceous ones have the same shape with a right-sided asymmetry; on the other hand, the most calcareous ones, like the slag, exhibit a left-sided asymmetry. The less amorphous samples have a symmetric hump, due to the sharp peaks which flatten the hump. The full width at half maximum (FWHM) for the silica fume and slag humps are half the size of the recycled glass.

The FWHM for the recycled glass is always larger than other samples. This shape leads us to suppose the presence of two contributions: one corresponds to the silica fume hump, and the second to the slag, at the characteristic positions (21 and 31°$2\theta_{Cu}$) respectively. The XRD-pattern of the recycled glass is very similar to the one obtained by Durdzinski *et al.* [81] and Schöler *et al.* [18]. Two different glass structures seem to be coexisting in the recycled glass. This hypothesis would be investigated using Raman spectroscopy.

## 3.3. Raman spectroscopy

3.3.1 Description of the whole Raman spectra

Raman microspectroscopy was used to study the glass part of all materials. Thus, in the following, the crystalline phase spectra are not shown but they correspond to the phases identified by XRD. A microscopic volume (circa 0.5-1 $\mu m^3$) was measured precisely in the glass part through the use of the microscope confocal mode.

Figure 3 shows representative spectra of some more relevant samples and a comparison with pure silica glass [82]. For clarity, only one spectrum from each material family is shown: S3 slag is representative for the three GGBF slags; Si FA2 is representative for silico-aluminous fly ashes; Ob is representative for obsidian and volcanic pozzolan; and for the two silica fumes, the sample SF2 gave the best signal-to-noise ratio. These spectra were completed by pure silica glass from the literature [11, 82], calcium fly ash Ca FA and the recycled glass Rec Gl. It is clear that the silica fume spectrum is very similar to the pure silica glass one. Despite the tiny spot size, the FA spectra show the D and G bands of coal at 1345 $cm^{-1}$ and 1606 $cm^{-1}$ [31, 32, 83, 84], logically found as fly ashes originating from coal-fired power plant. This organic part, always present in FA, contributes to the amorphous part of the material beside the mineral part. This organic part should not be confused with the burning effect that can be caused by focusing a powerful laser beam on the sample, as observed, for example, on obsidian. The volcanic pozzolan P2, not shown here, gives an unresolved large hump between 50 and 1200 $cm^{-1}$ with a maximum at 900 $cm^{-1}$, which could be attributed to the photoluminescence of the sample, and which prevents observation of the Raman peaks. For the calcareous fly ash (Ca FA), the frequency of crystalline phases makes it difficult to find the glass part of the material, requiring a larger number of spectra collections. In general, the glass parts are more reproducible than the spectra of the crystalline phases which can present different orientations, giving rise to polarization effects affecting the Raman spectra. The glass part cannot be confused with the crystalline part because of the full width at half maximum (FWHM) of the Raman bands. For crystalline phases, the band's FWHMs are below 20 $cm^{-1}$, while they are larger than 100 $cm^{-1}$ for glass. The wide bands of a glass illustrate its high dispersion in term of bond lengths and angles, illustrating a local and intermediate disorder. The water in the samples was not investigated.

The Raman spectra of the silica glass present peaks in the 50-1200 $cm^{-1}$ domain and are composed of four main regions: below 200 $cm^{-1}$ (very low-frequency region), from 200 to 600 $cm^{-1}$ (low-frequency region), from 600 to 800 $cm^{-1}$ (medium-frequency region), and from 800 to 1200 cm$^{-1}$ (high-frequency region). Glass spectra do not present peaks above 1200 cm$^{-1}$, the peaks found above this come from the other materials in the glass material (organic fraction for instance).

### 3.3.2. Siliceous glass fraction of SCMs: Boson peak (below 200 cm$^{-1}$)

Below 200 cm$^{-1}$, the Boson peaks are characteristic of glass materials. They appear on our spectra for a minimum $Al_2O_3$ content of 11 wt.% and when the glass is depolymerized (see 3.3.5). Their origin remains very controversial and there are several theories about it [11]. On our spectra, bands are observed at 91 cm$^{-1}$ for the slag, at 75 cm$^{-1}$ for the Ca FA, at 82 cm$^{-1}$ for the Si FA2, and at 83 cm$^{-1}$ for the obsidian. The more the glass is depolymerized, the greater the intensity.

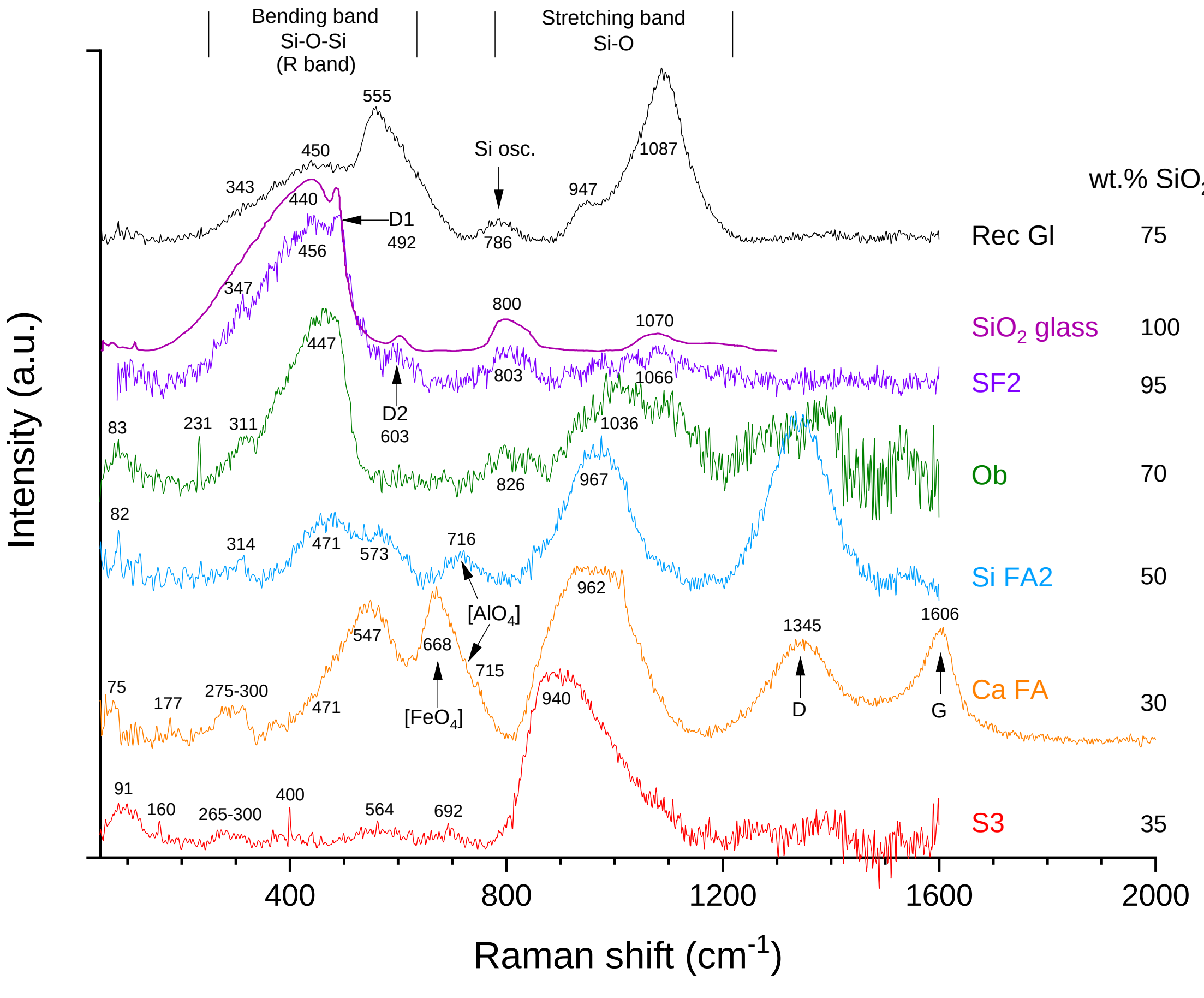

*Figure 3: Raman spectra of SCMs with various polymerization degrees (pink spectrum of $SiO_2$ glass is drawn after Mysen et al. [82]). The bulk content in wt.% $SiO_2$ is indicated on the right.*

### 3.3.3. Siliceous glass fraction of SCMs: stretching domain (800-1200 cm$^{-1}$)

The high-frequency region (800-1200 cm$^{-1}$) corresponds to the symmetric stretching bands of the Si-O bonds in $SiO_4$ tetrahedra (short range order, SRO). In Figure 3, the spectra were stacked from bottom to top, classified with the shift of the band to higher frequencies. The position toward higher frequencies goes hand in hand with an increase in the $SiO_2$ content in materials. It shows three classes of SCMs glasses with stretching bands found under the pure silica glass position (1070 $cm^{-1}$). The first comprises the positions for the slag (940 $cm^{-1}$), the calcareous FA (962 $cm^{-1}$), the siliceous FA2 (967 $cm^{-1}$) and for the obsidian (1036 $cm^{-1}$). Secondly, the silica fume stretching band is at similar position (1066 $cm^{-1}$) to silica glass. Finally, the recycled glass is particular, because the spectrum shows clearly two bands at 947 $cm^{-1}$ and at 1087 $cm^{-1}$, and these two bands are not as wide as the bands in other materials. The first one is at lower frequency than in pure silica glass and the second at higher frequency. These two band positions correspond exactly to the published data on the soda glass (15 wt.% $Na_2O$) spectrum [20].
The 800 $cm^{-1}$ band is assigned to stretching vibration due to the Si motion against its oxygen cage in the $SiO_4$ tetrahedron [38, 85]. The band is intense enough to be seen when the $SiO_2$ amount is above 70 wt.%, hence it appears on spectra only for obsidian (826 $cm^{-1}$), silica fume (803 $cm^{-1}$), pure silica glass (800 $cm^{-1}$), and recycled glass (786 $cm^{-1}$). The band shape is asymmetric for silica fume and pure silica glass (>90 wt.% of $SiO_2$) and is symmetric for obsidian and recycled glass.
The presence of sulfur in the glass in the Ca FA was investigated (7 wt.% of $SO_3$). A band at 995 $cm^{-1}$ is the characteristic band associated with sulfur in a glass [86], but is not found in Ca FA. All the content of $SO_3$ is only present in the crystalline phases of Ca FA, confirmed by reverse Bogue calculation (see discussion).

3.3.4. Siliceous glass fraction of SCMs: incorporation of Al and Fe (600 to 800 $cm^{-1}$)

In the medium-frequency region (600 to 800 $cm^{-1}$), bands are assigned to symmetric stretching vibration of $AlO_4$ unit (in $Q^2$) and $FeO_4$ unit in the glass as a network former. These bands are found in Si and Ca fly ashes as well as in slag to a minor degree. $FeO_4$ is incorporated only in the Ca FA glass (band at 668 $cm^{-1}$), while $AlO_4$ is also present in slag S3 (692 $cm^{-1}$), in Ca FA (715 $cm^{-1}$) and in Si FA2 (716 $cm^{-1}$). It should be noted that despite the high amount of $Al_2O_3$ in obsidian (13.5 in wt.%), it seems that $Al^{3+}$ is not necessarily as network former but could play a network modifier role.

3.3.5. Siliceous glass fraction of SCMs: degree of polymerization (800-1200 $cm^{-1}$)

Further structural knowledge can be extracted from the higher frequency bands, between 800-1200 $cm^{-1}$, which are associated with an increasing number of bridging oxygens (BO) by $SiO_4$ tetrahedron in the glass part, noted $Q^n$ (n being the number of bridging oxygens by tetrahedron) [85, 87]. These $Q^n$ are respectively named: $Q^0$ ($SiO_4^{4-}$ monomer), $Q^1$ ($Si_2O_7^{6-}$ dimer or a chain end group), $Q^2$ ($SiO_3^{2-}$ chain), $Q^3$ ($Si_2O_5^{2-}$ sheet or NBO in a framework structure) and $Q^4$ ($SiO_2$ three-dimensional network) [88]. Therefore, in order to determine the relative percentage of each $Q^n$ unit present in the glass part of SCMs, the stretching areas are smoothed, renormalized in this domain and finally decomposed by a curve-fitting procedure using 3 to 6 bands with Gaussian profiles as shown in Figure 4. Table 4 gives the assignment in terms of $Q^n$ for a band in a Raman shift domain. Slag and fly ashes (Ca FA and Si FA2) contain only low polymerized tetrahedra in $Q^0$, $Q^1$ and $Q^2$. Only slag and Ca FA contain $Q^0$ units. While Si FA has the same amount of $Q^1$ and $Q^2$, Ca FA contains many more $Q^2$ units. These compounds are poor in $SiO_2$ (glass former), but rich in glass modifiers (CaO, MgO). Obsidian seems to be special: although it has the same number of Q2 and Q3 units, the high content of Q1 units could be surprising. Al atoms do not seem to be incorporated in the glass structure (lack of $AlO_4$ specific stretching bands). The high value of the sum of glass modifiers ($Al_2O_3$+CaO+$K_2O$+$Na_2O$=24 wt.%) seems to explain the lower than expected polymerization. If the silica fume spectrum looks like pure silica glass, the decomposition highlights a difference in terms of $Q^n$ units. Pure silica glass contains only $Q^3$ and $Q^4$ units, and silica fume shows a broad range of $Q^n$ with only 75% of $Q^3$, $Q^4$, which would indicate that silica fume could have reacted due to its grain size (500 nm in average) confirmed by NMR [89, 90]. The recycled glass spectrum shows two distinct bands despite the high amount of $SiO_2$: 16% of $Q^2$ units and 84% of $Q^3$ units. The first band is positioned at the same position as the S3 slag band, and the second one at the same position as SF2 and pure silica glass. The recycled glass seems to contain, on the one hand, an $SiO_2$-rich domain, corresponding to high frequency stretching vibration, and, on the other, an $SiO_2$-depleted domain, containing mainly $Q^2$ units, which could be explained by a corresponding rich concentration area of Na modifiers.

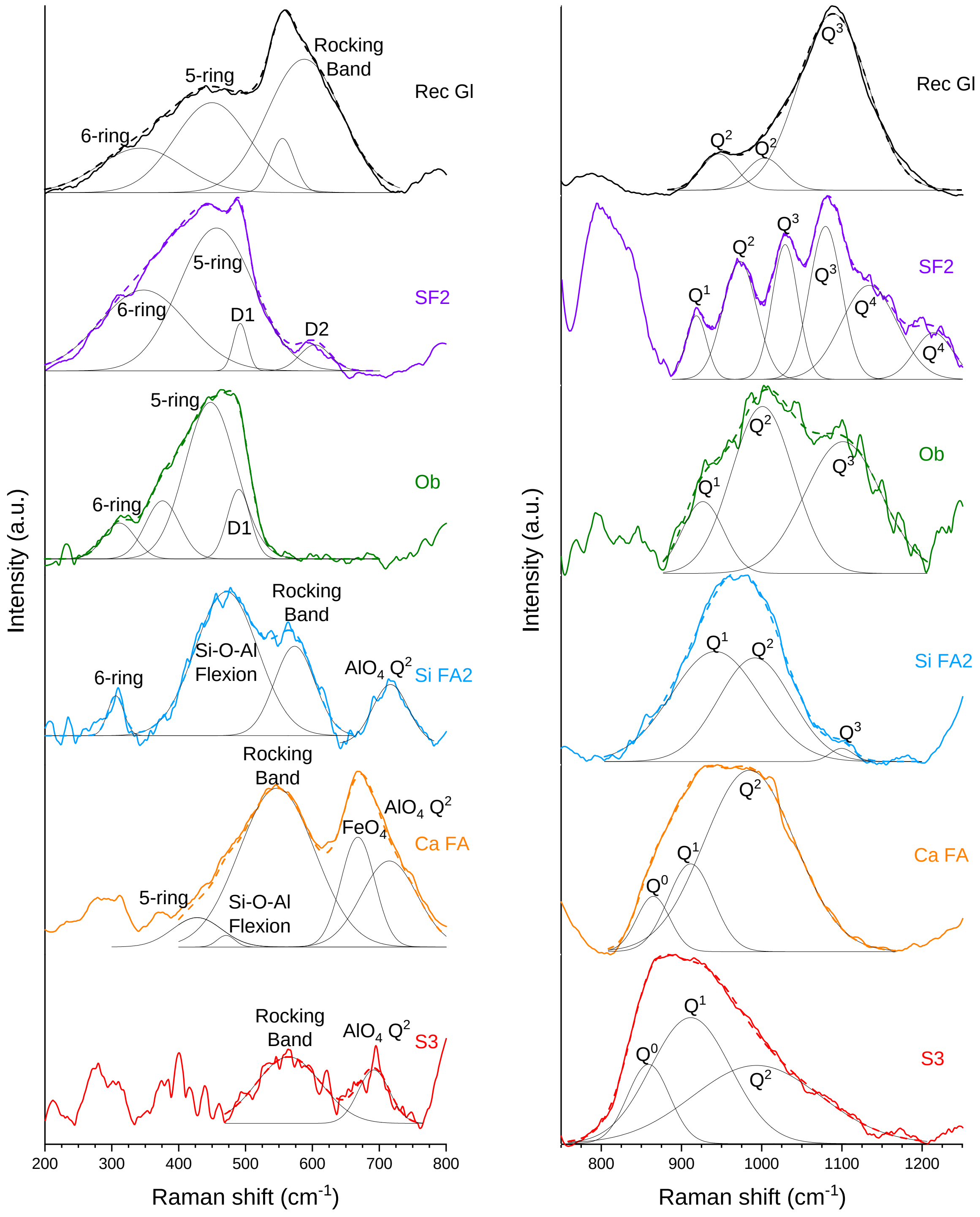

*Figure 4: Decomposition of the Raman spectra to determine the presence of each unit for bending motions (200-800 cm$^{-1}$) on the left, and for stretching motions (800-1200 cm$^{-1}$).*

### 3.3.6. Siliceous glass fraction of SCMs: relative intensities between stretching and bending bands

When comparing the entire spectrum of all SCMs (Figure 3), there is a variation of intensity between the stretching (800-1200 $cm^{-1}$) and the bending (200-600 $cm^{-1}$) domains. The relative intensities between stretching and bending bands depend on $SiO_2$ content and on polymerization degree. The bending band appears with an $SiO_2$ content of 70 wt.% and with the $Q^3$ unit presence in the glass. The silica fume spectrum has the same profile as pure $SiO_2$ glass, with a strong bending band and a very weak stretching signal. On the contrary, on the slag spectrum only the stretching band appears with a very weak bending signal. Finally, for fly ashes and obsidian spectra the two bands have equivalent intensities. The recycled glass is a soda glass, and the Na presence also implies a strong stretching band together with the bending band as already published [91]. The Ca-rich SCM samples (slag and Ca FA) are too poor in $SiO_2$ to show bending bands in their Raman spectra. The $Q^n$ units are not large enough to give a medium range order characteristic of silica glass structure [11].

*Table 4: Results of the decomposition of Raman bands in the bending and stretching domain of the glass part of some representative SCMs compared to pure $SiO_2$ glass [11, 82]: Raman shift in first line, area ratio for the stretching band in parenthesis, calculated Si-O-Si angle in n-ring in italics by Hehlen formula [92]. $Q^n$ domain references: [24-26, 38, 82, 85, 92-95].*

| | | **Raman shift domain ($cm^{-1}$)** | **S3** | **Ca FA** | **Si FA2** | **Ob** | **SF2** | **$SiO_2$ glass** | **Rec Gl** |
|---|---|---|---|---|---|---|---|---|---|
| **Si-O-Si vibration** | 6-ring in 3D | ~310 | | | 306<br>*155°* | 311<br>*155°* | | | |
| | Scissor | ~360 | | | | 376<br>*150°* | 347<br>*152°* | 370<br>*150°* | 343<br>*152°* |
| | Flexion 5-ring | ~440 | | 427<br>*145°* | | 447<br>*144°* | 456<br>*143°* | 449<br>*144°* | 450<br>*143°* |
| **Si-O-Al** | Flexion | | | 471 | 471 | | | | |
| **Si-O-Si vibration** | D1 4-ring | ~490 | | | | 491<br>*140°* | 492<br>*140°* | 488<br>*140°* | |
| | Rc Rocking band | | 564 | 547 | 573 | | | | 555<br>588 |
| | D2 3-ring | ~600 | | | | | 603<br>*130°* | 603<br>*130°* | |
| **$FeO_4$** | | 668 | | 668<br>(43) | | | | | |
| **$AlO_4$** | $Q^2$ | 715 | 692<br>(100) | 715<br>(57) | 716<br>(100) | | | | |
| **$SiO_4$ vibration** | $Q^0$ | 800-890 | 859<br>(14) | 865<br>(8) | | | | | |
| | $Q^1$ | 900-920 | 911<br>(43) | 909<br>(18) | 942<br>(55) | 929<br>(13) | 918<br>(6) | | |
| | $Q^2$ | 950-1030 | 996<br>(43) | 985<br>(74) | 992<br>(43) | 1003<br>(44) | 973<br>(19) | | 947<br>(16) |
| | $Q^3$ | 1030-1100 | | | 1101<br>(2) | 1103<br>(43) | 1080<br>(41) | 1070<br>(55) | 1087<br>(84) |
| | $Q^4$ | 1100-1200 | | | | | 1137<br>(34) | 1180<br>(45) | |

3.3.7. Siliceous glass fraction of SCMs: bending bands (200 and 600 $cm^{-1}$)

In the low-frequency region (Figure 3), between 200 and 600 $cm^{-1}$, the wide asymmetric band corresponds to the bending vibration (flexion) of the inter-tetrahedral of Si-O-Si, and highlights the connectivity of the silicate networks in the form of n-rings (n-ring = n-membered ring) of different sizes. In silicate glass, rings are defined in terms of T-O links (T=Si, Al, Fe or Ti here) forming rings [94]. The ring size is a measure of the medium range order (MRO). Band positions give the size of the rings. For pure silica glass the 449 $cm^{-1}$ band corresponds to a 5-ring and 370 $cm^{-1}$ to a 6-ring. The so-called breathing bands D1 at 488 $cm^{-1}$ and D2 at 603 $cm^{-1}$ [38, 94] correspond to a 4-ring and a 3-ring respectively. D1 (492 $cm^{-1}$) and D2 (603 $cm^{-1}$) bands are at the same position for silica fume. D1 together with D2 are observed only for high $SiO_2$ content glasses (SF2 and Ob). The flexion bands between pure silica glass and silica fume are very close in terms of shape but not in terms of full width at half maximum (FWHM). The FWHM of the SF2 bending band is thinner (difference of 30 $cm^{-1}$) than that of pure silica glass. Obsidian contains 4-ring (D1 at 491 $cm^{-1}$), 5-ring (band at 447 $cm^{-1}$), and 6-ring (band at 376 $cm^{-1}$ and 311 $cm^{-1}$). 471 $cm^{-1}$ is a common band found in Ca FA and Si FA and seems to correspond to an Si-O-Al flexion vibration in a 5-ring network. Rocking vibration is observed at around 550-570 $cm^{-1}$: 547 $cm^{-1}$ is assigned to Si-O-Fe in Ca FA, whereas 573 $cm^{-1}$ in Si FA2, and 564 $cm^{-1}$ in S3 to Si-O-Al (ionic radius: $Fe^{3+}$=0.67 Å and $Al^{3+}$=0.53 Å).

The Si-O-Si angles between tetrahedra were calculated using the Hehlen formula [92] based on the maximum position of the bending band compared to the pure silica acting as the reference (426 $cm^{-1}$ for Si-O-Si angle=145.5°). The Si-O-Si angles calculated after the curve-fitting procedure (Figure 4) are given in Table 4. The higher the size of a ring, the higher is the Si-O-Si angle in this ring. A shift of the Si-O-Si bending band toward a higher frequency corresponds to a lower angle [96]. The 471 $cm^{-1}$ band should then correspond to Si-O-Al bending motion with a smaller angle than Si-O-Si [97] ($Si^{4+}$=0.41 Å - $Al^{3+}$=0.53 Å).

3.3.8. Bimodality of the recycled glass

Recycled glass is particular, with two bending bands (450 and 555-588 $cm^{-1}$) coexisting with two stretching bands (947 and 1087 $cm^{-1}$). These two couples of bending and stretching bands are the criterion to have a bimodal glass [36, 91]. The first domain is characterized by the flexion band at 450 $cm^{-1}$, which corresponds to the stretching band at 1087 $cm^{-1}$, meaning that 84% of the recycled glass is made of 5-ring of $Q^3$ tetrahedra assembled with an Si-O-Si angle of 143°. The remaining 16% are related to the two rocking bands at 555 $cm^{-1}$ and 588 $cm^{-1}$ together with the 947 $cm^{-1}$ stretching band. They are constituted of 3-ring of $Q^2$ tetrahedra with Si-O-Si angle of 134° and 132° respectively, because a small Si-O-Si angle is linked to long Si-O bond length [98]. In this domain the Na concentration should be high enough to cut the 5-ring generating smaller rings with smaller angles and weakening the bonding force. This could be illustrated by the modified random network (MRN) of Greaves [14]. The recycled glass is inhomogeneous and contains two intimately interlocked domains: one with a silica glass-type structure (1087 $cm^{-1}$) and one with a slag glass-type structure (947 $cm^{-1}$). The two XRD hump positions of recycled glass (Figure 2) could now be explained with a slag glass-type structure (CaO-rich, at 30.6°$2\theta_{Cu}$), and a silica glass-type structure ($SiO_2$-rich, at 22.8°$2\theta_{Cu}$). These two domains will be designated for the rest of discussion by “Ca Rec Gl” and “Si Rec Gl” respectively.

## 4. DISCUSSION

The cross-referencing of XRD and Raman spectroscopy data appears to be relevant. These two technical analyses give access to different structural information on the materials. Powder XRD is the result of X-Ray scattering from crystallized (tens of thousands of randomly oriented crystallites) and glass parts of the materials. Conversely, Raman spectroscopy is conducted on a very focused zone on the glass part of materials, and this technique is very local at the atomic level.

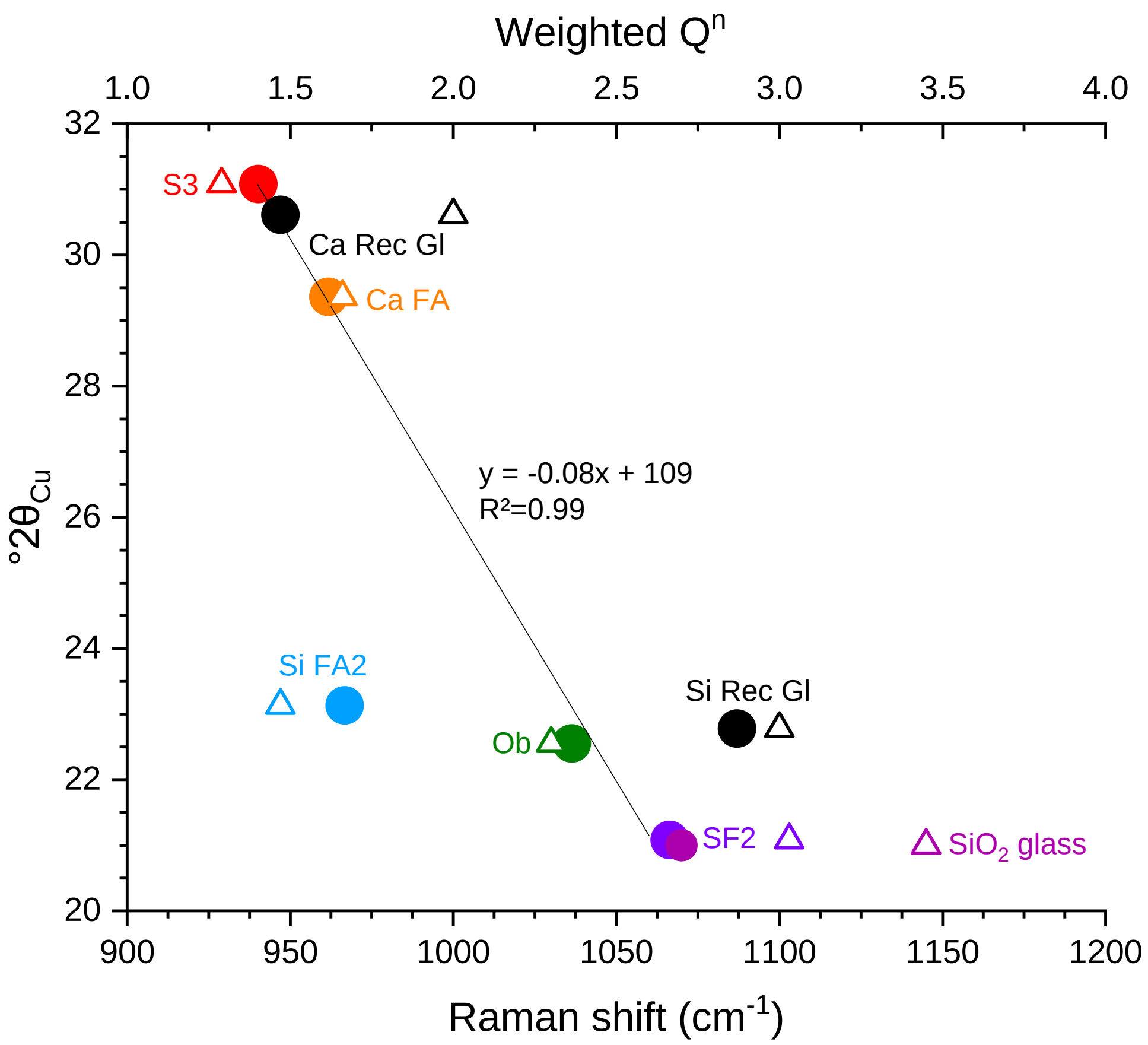

*Figure 5: Maximum hump angular position of SCMs as a function of the stretching band maximum (below, in circle) and the weighted $Q^n$ (above, in triangle) for each SCM.*

Maximum hump angular positions of SCMs as a function of the Raman $SiO_4$ stretching band maxima and also as a function of weighted $Q^n$ for each SCM are illustrated in Figure 5. If Si FA2 and Si Rec Gl are not considered, a strong correlation ($R^2$=0.99) was observed between the hump position and the stretching band position. Raman spectrum acquisition takes just a few minutes, whereas XRD pattern acquisition takes at least half a day to highlight the hump sufficiently. The Raman position is quickly able to give the XRD hump position except for silico-aluminous fly ash. Two trends emerge from this diagram: on the one hand, samples with the high rate of depolymerization ($Q^1$-$Q^2$) are linked to a hump around 30°2$\theta_{Cu}$, but on the other hand samples with a low rate of depolymerization ($Q^2$-$Q^3$) are characterized by a hump around 22°2$\theta_{Cu}$. Hump position can be seen as a signature of depolymerization. With regard to recycled glass, it is noticeable that it is split into two distinct domains, one corresponding to a highly depolymerized region (low Raman shift and low $Q^n$), and the other corresponding to a highly polymerized region (high Raman shift and high $Q^n$). From Figure 5, the Raman shift of silico-aluminous fly ash indicates strong depolymerization, and its Raman shift position could suggest a hump position at a higher value, around 28°2$\theta_{Cu}$, rather than at 23°2$\theta_{Cu}$, which corresponds to Si-rich glasses.

Links between the results of XRD and XRF data were also investigated to find out if available relations could exist for the thirteen SCMs with different levels of approximations, firstly using the bulk chemical composition of the materials (data of Table 2), and secondly going further, using the calculated chemical composition of the glasses, after subtraction of the oxides from crystalline phases by the reverse Bogue method (data of Table 5).

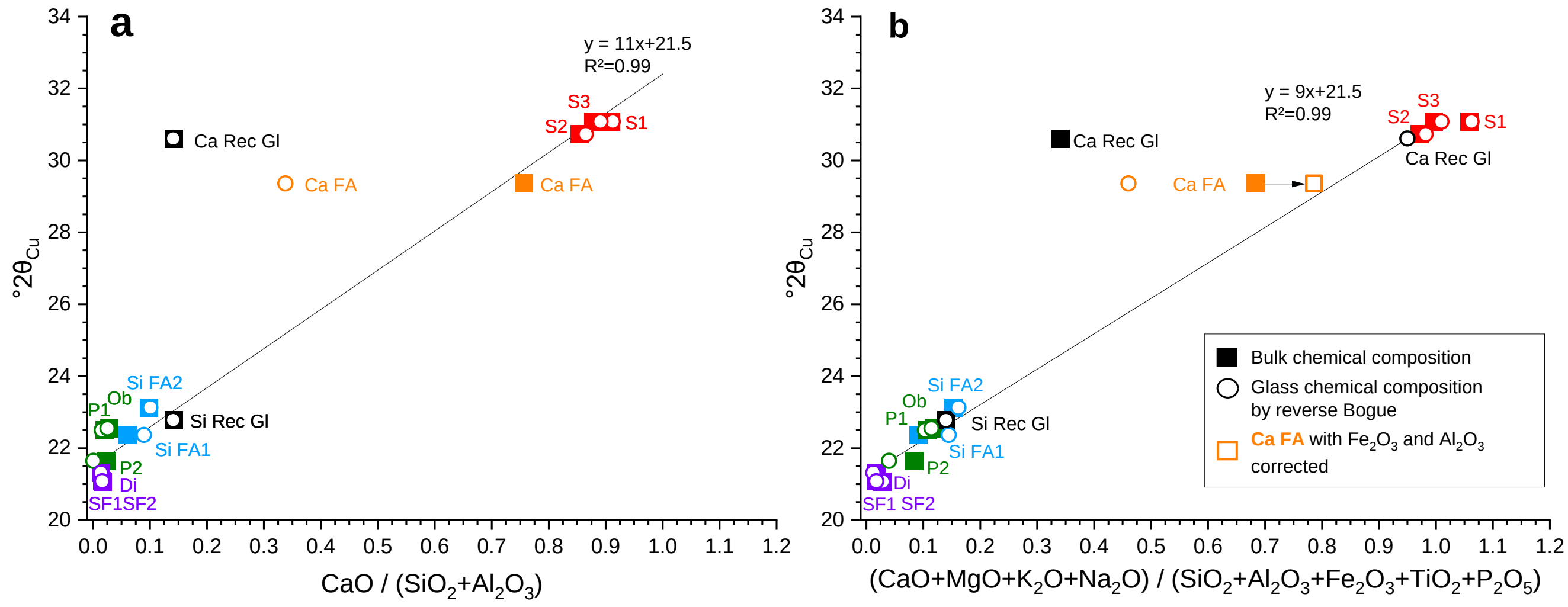

*Figure 6: Maximum hump angular position of SCMs of this study* ***a)*** *as a function of the ratio* $CaO/(SiO_2+Al_2O_3)$ ***b)*** *as a function of the ratio modifiers/formers* $((CaO+MgO+K_2O+Na_2O)/(SiO_2+Al_2O_3+Fe_2O_3+TiO_2+P_2O_5))$*. On the two figures solid squares correspond to bulk XRF composition and empty circles to the calculated glass composition. For recycled*

*glass, the two contributions Ca Rec Gl and Si Rec Gl are reported. In figure b, the empty squares for Ca FA corresponds to a correction of only $Al_2O_3$ and $Fe_2O_3$ of crystalline phases of Ca FA. The better fit is clearly marked by an arrow.*

First, a plot of the maximum hump position from XRD is drawn in Figure 6a, as a function of the main oxide ratio $CaO/(SiO_2+Al_2O_3)$ from the XRF bulk chemical composition of the materials. The relation using the bulk chemical composition already gives an excellent fit ($R^2$=0.99). Again, the correlation was made excluding the Ca Rec Gl contribution of recycled glass. This good correlation can be surprising for SCMs containing a lot of crystalline phases, such as Ca FA. This relation can be used to assess the range of order of the three major chemical oxides content, when the maximum hump position is known. Conversely, knowing only the quantity of the three major oxides means that the hump position can be predicted.

To go further, the same correlation was made using the calculated chemical composition of the glasses (Figure 6a). For Si FA1 only, the correlation is better when the glass chemical composition is used. It is not better for silica fumes SF1 and SF2, Si FA2 and the silica glass-type structure of recycled glass Si Rec Gl. However, for P2, Ca FA, and the three slags S1, S2 and S3, the correlation is better with the bulk chemical composition ratio. However, the calculation based on the glass composition provides a better analysis of the hump position for fly ashes. For the recycled glass, the slag glass-type structure (Ca Rec Gl) is far off the correlation because the significant $Na_2O$ content (12 wt.%) is not taken into account in the ratio $CaO/(SiO_2+Al_2O_3)$.

In Figure 6b, to take into account more precisely the glass structure in terms of modifiers and formers, the hump maximum positions are plotted against the ratios of modifiers to formers ($(CaO+Na_2O+MgO+K_2O)/(SiO_2+Al_2O_3+Fe_2O_3+TiO_2+P_2O_5)$) expressed in weight percent. The correlation is firstly made with the bulk chemical composition represented by solid squares in Figure 6b. The correlation between hump position and modifiers to formers ratio is more precise than the $CaO/(SiO_2+Al_2O_3)$ ratio as it takes into account more oxides. $SiO_2$-rich SCMs show a better alignment comparing square positions (bulk composition) between Figure 6b and Figure 6a, except for P2. Two slags out of three are exactly correlated: only S1 does not perfectly fit with the global relation. Its high content of MgO (7 wt.%), with a ratio $MgO/Al_2O_3$=0.56, could explain the poorer match with the correlation. With the high Al content, Mg which is classically a glass network modifier, can become a glass network former [99]. For Ca FA, the correlation matches less when the hump position is expressed versus the $CaO/(SiO_2+Al_2O_3)$ ratio because $Al_2O_3$ and $Fe_2O_3$ are incorporated in the glass. As $Al_2O_3$ and $Fe_2O_3$ are also present in crystalline phases (reported in Table 3), their contents in these phases have been corrected (Table 5). The new position of Ca FA is marked by an arrow on Figure 6b and fits better with the correlation, but is still equivalent to the hump position expressed versus $CaO/(SiO_2+Al_2O_3)$ (empty square Figure 6b versus solid square Figure 6a).

As the recycled glass is bimodal, not all oxides have been taken into account in the $(CaO+Na_2O+MgO+K_2O)/(SiO_2+Al_2O_3+Fe_2O_3+TiO_2+P_2O_5)$ ratio. For the Si Rec Gl part, the representative point was plotted considering only CaO in the numerator $(CaO)/(SiO_2+Al_2O_3+Fe_2O_3+TiO_2+P_2O_5)$, and for the Ca Rec Gl, only $SiO_2$ was considered in the denominator $(CaO+Na_2O+MgO+K_2O)/(SiO_2)$. The $SiO_2$-rich domain (Si Rec Gl) fits well. For the CaO-rich domain, taking into account $Na_2O$, $K_2O$ and MgO in addition to CaO remains far from the correlation because the $SiO_2$ content in this domain is unknown. To estimate the $SiO_2$ distribution in the two domains, the correlation of the hump position with the ratio of modifiers to formers is used. The distribution of the chemical composition of $SiO_2$ and CaO in the two domains is adjusted to fit these two data with the linear regression (black empty circle Figure 6b). After calculation, 76% of $SiO_2$ present in the recycled glass can be assigned to the $SiO_2$-rich domain (Si Rec Gl), and 24% to the CaO-rich domain (Ca Rec Gl). This fits with the distribution of Si-O stretching found by Raman spectroscopy, with 84% $Q^3$ for Si Rec Gl and 16% $Q^2$ for Ca Rec Gl. It can be estimated that the recycled glass consists of 80% of silica glass-type structure and 20% of slag glass-type structure. Further corrections are still necessary to determine with precision the distribution of the two domains in the recycled glass.

The points calculated from the ratio of modifiers to formers based on the glass chemical composition are at the same position for SF1, SF2, Di, Ob, P1 and Si Rec Gl as those calculated using the bulk chemical composition and were already well aligned with the linear correlation. The Si FA1, Si FA2 and P2 positions are better aligned using the calculated glass chemical composition, particularly P2. Conversely, for S1, S2, S3 and Ca FA the bulk chemical composition allows a better fit. For Ca FA, the correction with only $Al_2O_3$ and $Fe_2O_3$ fits better with the line than with the reverse Bogue correction (with the glass chemical composition). Globally the data are more aligned using the bulk chemical composition than using the calculated glass composition except siliceous fly ashes and the pozzolan P2.

In Figure 7, the glass compositions calculated by reverse Bogue are placed in the ternary phase diagram. This highlights that for fly ashes, strong corrections (see arrows) appear due to their low amorphous content.

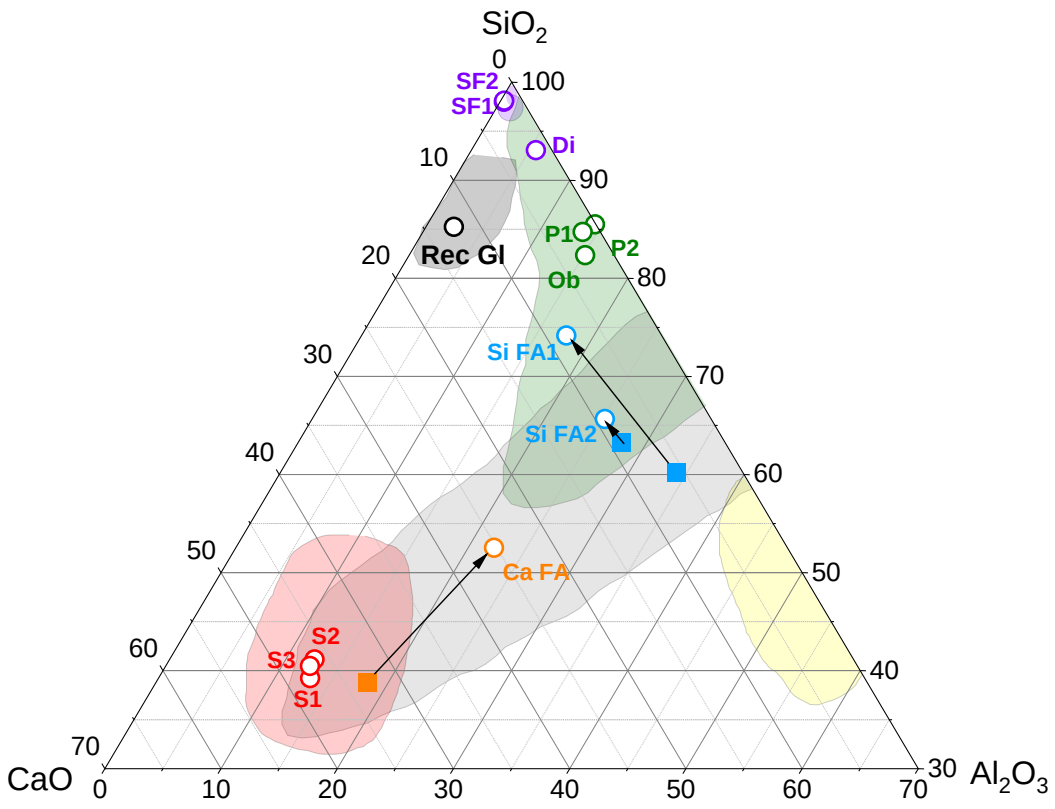

*Figure 7: Ternary phase diagram CaO-$SiO_2$-$Al_2O_3$ (wt.%) with the analyzed samples (filled square) and with glass composition of fly ashes calculated by reverse Bogue (empty circle). For high amorphous SCMs, bulk and glass chemical composition overlap and only glass chemical compositions are apparent.*

*Table 5: Chemical composition of the crystalline phases calculated by reverse Bogue, and of glass in bold (wt.%).*

| Sample | LOI | $SiO_2$ | $Al_2O_3$ | $Fe_2O_3$ | CaO | MgO | $SO_3$ | $K_2O$ | $Na_2O$ | $P_2O_5$ | $TiO_2$ | MnO | SrO |
|---|---|---|---|---|---|---|---|---|---|---|---|---|---|
| S1 | 0.30 | 0.72 | 0.17 | | 0.63 | | 0.07 | | 0.11 | | | | |
| | **0** | **34.68** | **11.53** | **0.3** | **42.17** | **7.2** | **2.00** | **0.4** | **0.22** | | **0.54** | **0.11** | **0.06** |
| S2 | 0.24 | 1.10 | 0.27 | | 0.63 | | 0.09 | | 0.17 | | | | |
| | **0** | **36.7** | **11.13** | **0.4** | **41.37** | **6.4** | **1.28** | **0.3** | **0.04** | **0.01** | **0.76** | **0.38** | **0.04** |
| S3 | | 0.69 | 0.10 | | 0.10 | 0.01 | 0.12 | | | | | | 0.18 |
| | **0** | **36.51** | **11.2** | **0.5** | **42.5** | **6.29** | **1.38** | **0.4** | **0.25** | **0.01** | **0.75** | **0.18** | **0** |
| Ca FA | 0.49 | 15.51 | 7.93 | 4.07 | 26.85 | 2 | 6.80 | | 0.06 | | | | |
| | **0.72** | **15.29** | **6.47** | **4.83** | **7.35** | **0.90** | **0.20** | **0.3** | **0.09** | **0.11** | **0.76** | **0.18** | **0.07** |
| Si FA1 | 0.43 | 18.06 | 20.89 | 1.40 | 1.21 | | | | | | | | |
| | **10.17** | **32.44** | **7.71** | **3.70** | **3.59** | **1.2** | **0.28** | **1.5** | **0.38** | **0.77** | **1.48** | **0.04** | **0.17** |
| Si FA2 | | 11.91 | 7.39 | 1.90 | 1.74 | 1.00 | 0.26 | | | | | | |
| | **4.96** | **37.79** | **14.51** | **5.10** | **5.26** | **1.50** | **3.74** | **1.7** | **1.09** | **0.83** | **0.88** | **0.05** | **0.31** |
| P2 | | 19.8 | 4.07 | | 1.86 | 0.56 | 1.02 | 0.68 | 1.61 | | | | |
| | **16.22** | **51.10** | **8.63** | **1.1** | **0.04** | **0** | **0** | **3.22** | **0** | **0.01** | **0.12** | **0.05** | |
| P1 | | 1.31 | 0.3 | | 0.40 | 0.04 | 0.35 | 0.11 | | | | | |
| | **5.11** | **72.69** | **11.8** | **1.2** | **1.30** | **0** | **0** | **4.09** | **3.46** | **0.01** | **0.13** | **0.04** | |
| Rec Gl | 0.09 | 0.59 | 0.07 | | 0.15 | | | | | | | | |
| | **0.36** | **71.51** | **2.03** | **0.4** | **10.35** | **1.3** | **0.05** | **0.6** | **12.15** | **0.02** | **0.05** | **0.03** | **0.04** |
| Ob | | 0.41 | 0.15 | 0.40 | 0.31 | 0.04 | 0.36 | 0.13 | | | | | |
| | **0.66** | **72.59** | **13.35** | **0.60** | **2.19** | **0.05** | **0** | **3.67** | **4.01** | **0.02** | **0.15** | **0.09** | **0.01** |
| Di | 0.04 | 1.38 | 0.20 | 2.89 | 0.05 | 0.51 | | | | | | | |
| | **19.96** | **78.52** | **4.70** | **0** | **1.15** | **0** | | **0.08** | **0.1** | **0.15** | **0.17** | **0.07** | **0.08** |
| SF1 | | 2.5 | | 0.03 | | | | | | | | | |
| | **6.07** | **88.4** | **0.3** | **0.01** | **1.5** | **0.2** | **0.02** | **0.6** | **0.24** | **0.08** | **0.01** | **0.01** | |
| SF2 | | | | | | | | | | | | | |
| | **0.4** | **93.5** | **0.3** | **0.04** | **1.5** | | | | **0.15** | | **0.05** | | |

# 5. CONCLUSION

This paper is dedicated to the characterization of thirteen industrially used natural and artificial SCMs, covering a wide range of chemical compositions. They were analyzed by XRF, XRD and Raman microspectroscopy to investigate their glass structure. Raman analysis is a powerful tool for acquiring better knowledge of the glass structure, in order to characterize the degree of glass polymerization in terms of $Q^n$ units, estimate the connectivity of the silicate networks in terms of n-rings, and highlight the incorporation of minor elements into the glass network.

Combining XRD measurement with Raman spectroscopy shows that the XRD hump position of any given SCM is strongly correlated with its stretching band Raman shift, thereby providing a good indicator of the glass depolymerization degree. In the case of recycled glass which remains a potential SCM, the XRD diffusion hump is twice as wide as the other glasses (silicate and calcium aluminosilicate glass). This large hump can be fitted with two contributions: on the one hand the GGBS position and on the other the silica glass position. The Raman analysis confirms the existence of the bimodality of the glass, with a high depolymerized zone estimated at 20 wt.% and a very low depolymerized zone estimated at 80 wt.%.

Moreover, the XRD hump position shifts with the value of the $CaO/(SiO_2+Al_2O_3)$ ratio, according to the XRF bulk chemical composition of the thirteen materials. This correlation is surprisingly quite adequate, in particular for the low amorphous content samples (fly ashes), resulting from XRF bulk chemical composition and regardless of the glass chemical composition estimated by reverse Bogue calculation. For the siliceous samples, the correlation between the XRD hump position and the $(CaO+Na_2O+MgO+K_2O)/(SiO_2+Al_2O_3+Fe_2O_3+TiO_2+P_2O_5)$ ratio is slightly improved by considering the bulk as well as the glass composition.

## ACKNOWLEDGMENTS

This work was supported by France Ciment. Enrique Garcia-Caurel (PICM, Ecole Polytechnique) is warmly thanked for his advice for Raman data processing. The authors express their thanks for the support of the X-Ray crystallography facility, DIFFRAX, in Ecole Polytechnique. The authors are very grateful to the referees for their careful reading of the manuscript and their suggestions which improve the manuscript.